\documentclass[12pt,a4paper]{article}

\usepackage{a4wide}
\usepackage{amsmath}
\usepackage{bm}
\usepackage{amssymb}
\usepackage{graphicx}
\usepackage{cite}
\usepackage{epic}

\newcommand{\Op}{\mathcal{O}}
\newcommand{\HS}{S}

\newcommand{\M}{j}

\newcommand{\zt}{\zeta_3}

\newcommand{\zf}{\zeta_5}

\newcommand{\cN}{{\cal N}}

\begin{document}

\thispagestyle{empty}

\begin{flushright}\footnotesize
\texttt{HU-Mathematik-2014-08}\\
\texttt{HU-EP-14/16}\\
\vspace{0.5cm}
\end{flushright}
\setcounter{footnote}{0}

\begin{center}
{\Large{\bf
Non-planar anomalous dimension of twist-2 operators:\\[2mm]
higher moments at four loops
}} \vspace{15mm}

{\sc
V.~N.~Velizhanin}\\[5mm]

{\it Institut f{\"u}r  Mathematik und Institut f{\"u}r Physik\\
Humboldt-Universit{\"a}t zu Berlin\\
IRIS Adlershof, Zum Gro\ss{}en Windkanal 6\\
12489 Berlin, Germany\\
and
}\\
{\it Theoretical Physics Division\\
Petersburg Nuclear Physics Institute\\
Orlova Roscha, Gatchina\\
188300 St.~Petersburg, Russia}\\[5mm]

\textbf{Abstract}\\[2mm]
\end{center}

\noindent{
We compute the non-planar contribution to the anomalous dimension of the eight moment of the twist-2 operators in $\cN=4$ supersymmetric Yang-Mills theory at four loops. This result was obtained from the calculations of some elements of the anomalous dimension matrix for twist-2 operators with the help of a relations between its elements and eigenvalues. The properties of the obtained result and a possible relations with a planar results are discussed.

}
\newpage

\setcounter{page}{1}

\section{Introduction}
The great interest in the calculations of anomalous dimensions of the composite operators comes from the investigations of integrability in the framework of AdS/CFT-correspondence~\cite{Maldacena:1997re,Gubser:1998bc,Witten:1998qj}. In the planar limit there are different calculations~\cite{Anselmi:1998ms,
LN4,
Bianchi:2000hn,
Arutyunov:2001mh,
Dolan:2001tt,
Kotikov:2003fb,
Kotikov:2004er,
Eden:2004ua,
Bern:2006ew,
Kotikov:2007cy,
Fiamberti:2007rj,
Fiamberti:2008sh,
Velizhanin:2008jd,
Velizhanin:2008pc} for the test of Asymptotic Bethe Ansatz~\cite{Minahan:2002ve,
Beisert:2003tq,
Beisert:2003yb,
Beisert:2004hm,
Staudacher:2004tk,
Beisert:2005fw,
Beisert:2006ez,
Beisert:2007hz,
Bena:2003wd,
Dolan:2003uh,
Kazakov:2004qf,
Beisert:2005bm,Beisert:2005di,Arutyunov:2004vx,Beisert:2005cw,Janik:2006dc,Hernandez:2006tk,Arutyunov:2006iu,Beisert:2006ib} as well as for a spectral equations for the finite length operators~\cite{Arutyunov:2009zu,
Gromov:2009tv,Bombardelli:2009ns,Gromov:2009bc,Arutyunov:2009ur,Arutyunov:2009ax}. The calculations are performed perturbatively up to four/five loops~\cite{Anselmi:1998ms} for twist-2/twist-3 operators and up to one loop more with the using of generalized L\"usher corrections~\cite{Luscher:1985dn,Luscher:1986pf} for the operators with arbitrary Lorentz spin~\cite{Bajnok:2008bm,
Bajnok:2008qj,
Beccaria:2009eq,
Bajnok:2009vm,
Lukowski:2009ce,
Velizhanin:2010cm}.

For a non-planar case there are much less results (see Ref.~\cite{Kristjansen:2010kg}). However, for the {\it direct diagrammatic} calculations there is no a large difference between the planar and non-planar computations. Some time ago we calculated the anomalous dimension of the Konishi operator at fourth order in $\cN=4$ SYM theory~\cite{Velizhanin:2009gv}, where non-planar contribution appears for the first time for the twist-2 operators. The result was rather surprising, since it contains only $\zf$ contribution without $\zt$ and rational parts:
\begin{equation}
\gamma^{\mathrm {4-loop,\,np}}_{\mathrm{Konishi}}= -\frac{17280}{N_c^2}\
\zf\,g^8,\qquad g^2\ =\ \frac{g^2_{YM}\,N_c}{(4\,\pi)^2}\,. \label{resnpK}
\end{equation}
The calculations of the $\zf$ contribution to the first  three even moments allowed us to assume the following general form of non-planar contribution to the anomalous dimension of twist-2 operators with arbitrary Lorentz spin:
\begin{eqnarray}
\gamma_{{\mathrm{uni,np}}}(j)&\ =&\ -640\; S_1^2(j-2)\,\frac{12}{N_c^2}\
\zf\,g^8\,+\,...,\qquad S_1(j)=\sum_{i=1}^j\frac{1}{i}\,. \label{resnpuad}
\end{eqnarray}
However, this result is in contradiction with the usual large-$j$ behavior of anomalous dimension, which is expected to be proportional  $\ln j$ (see Ref.\cite{Alday:2007mf}), while from Eq.(\ref{resnpuad}) we obtain $(\ln j)^2$.

Then we calculate the next, second moment for the non-planar contribution to the anomalous dimensions of twist-2 operators at fourth order~\cite{Velizhanin:2010ey}. These results allowed to obtain the expressions for the fourth and sixth moments of the universal anomalous dimension of the twist-2 operators  using the fact, that the non-planar contribution appears at fourth-loop order for the first time and in this case we can use a superconformal symmetry of operators:
\begin{eqnarray}
\gamma_{\mathrm {uni,np}}(4)&= &
- 360\ \zf\, \frac{48\,g^8}{N_c^2}+\mathcal{O}\!\left(g^{10}\right)\,,
\label{guniM4}\\[2mm]
\gamma_{\mathrm {uni,np}}(6)&= &
\frac{25}{9}\bigg(21 + 70 \ \zt - 250 \ \zf \bigg)\,  \frac{48\,g^8}{N_c^2}+\mathcal{O}\!\left(g^{10}\right)\,.
\label{guniM6}
\end{eqnarray}
Unfortunately, the obtained information were not enough to reconstruct a general result for the non-planar contribution to the four-loop anomalous dimension of twist-2 operators for the arbitrary Lorenz spin of operators. So, we will need to find some more information.

In this paper we present the result of calculations of the non-planar (or color subleading) contribution for the next, eight moment of the four-loop universal anomalous dimension of twist-2 operators in $\cN=4$ SYM theory.
We discuss also a some general properties of the obtained result.

\section{Calculations and Result}
The calculations were performed in the same way, as in our previous works~\cite{Velizhanin:2008pc,Velizhanin:2010ey}. We consider the following ``QCD-like'' colour and $SU(4)$ singlet local Wilson twist-2 operators:
\begin{eqnarray}
\mathcal{O}_{\mu _{1},...,\mu _{\M}}^{g} &=&\hat{S} G_{\rho \mu_{1}}^{a}{\mathcal
D}_{\mu _{2}} {\mathcal D}_{\mu _{3}}...{\mathcal D}_{\mu _{\M-1}}G_{\rho \mu
_{\M}}^a\,,
\label{ggs}\\
\mathcal{O}_{\mu _{1},...,\mu _{\M}}^{\lambda } &=&\hat{S}
\bar{\lambda}_{i}^{a}\gamma _{\mu _{1}}
{\mathcal D}_{\mu _{2}}...{\mathcal D}_{\mu _{\M}}\lambda ^{a\;i}\,, \label{qqs}\\
\mathcal{O}_{\mu _{1},...,\mu _{\M}}^{\phi } &=&\hat{S}
\bar{\phi}_{r}^{a}{\mathcal D}_{\mu _{1}} {\mathcal D}_{\mu _{2}}...{\mathcal
D}_{\mu _{\M}}\phi _{r}^{a}\,,\label{phphs}
\end{eqnarray}
where ${\mathcal D}_{\mu_i }$ are covariant derivatives. The spinors $\lambda_{i}$ and field tensor $G_{\rho \mu }$ describe gauginos and gauge fields, respectively, and $\phi _{r}$ are the complex scalar fields appearing in the ${\mathcal N}=4$ SYM theory. Indices $i=1,\cdots ,4$ and $r=1,\cdots ,3$ refer to $SU(4)$ and $SO(6)\simeq SU(4)$ groups of inner symmetry, respectively. The symbol $\hat{S}$ implies a symmetrization of each tensor in the Lorentz indices $\mu_{1},...,\mu _{\M}$ and a subtraction of its traces. These operators mix with each other under renormalization and the eigenvalues of the matrix are expressed through the one function with the shifted argument, usually called universal anomalous dimension
(see Ref.~\cite{Kotikov:2002ab,Kotikov:2003fb} for details)
\begin{equation}
\gamma_{\mathrm {uni}}(\M)\ =\ \sum_{n=0}^{\infty}\gamma_{\mathrm {uni}}^{(n)}(\M)\,g^{2(n+1)}\,.
\end{equation}

The matrix, which diagonalize the matrix of anomalous dimension, can be used for the construction of the multiplicatively renormalizable operators. These operators will mix under renormalization and the anomalous dimension matrix will have triangle form in the higher-loop orders. However, the diagonal elements will give the universal anomalous dimension with shifted argument~\cite{Kotikov:2003fb}. These multiplicatively renormalizable operators have the following general form:
\begin{eqnarray}
\Op^{T_j}_{\mu_1,\ldots,\mu_j} &\ =\ & \Op^g_{\mu_1,\ldots,\mu_j} +
\Op^\lambda_{\mu_1,\ldots,\mu_j} + \Op^\phi_{\mu_1,\ldots,\mu_j}\,,
\label{mrop1j}\\[1mm]
\Op^{\Sigma_j}_{\mu_1,\ldots,\mu_j} &\ =\ & -\, 2\,(j-1)\,\Op^g_{\mu_1,\ldots,\mu_j} +
\Op^\lambda_{\mu_1,\ldots,\mu_j} + \frac{2}{3}\,(j+1)\,\Op^\phi_{\mu_1,\ldots,\mu_j}\,,
\label{mrop2j}\\[1mm]
\Op^{\Xi_j}_{\mu_1,\ldots,\mu_j} &\ =\ & -\, \frac{j-1}{j+2}\,\Op^g_{\mu_1,\ldots,\mu_j}
+ \Op^\lambda_{\mu_1,\ldots,\mu_j} -
\frac{j+1}{j}\,\Op^\phi_{\mu_1,\ldots,\mu_j}\,.
\label{mrop3j}
\end{eqnarray}
The combination of the anomalous dimensions for these operators sandwiched between some (gauge field, gaugino or scalar) state will give the universal anomalous dimension with an appropriate shifted argument and with a corresponding normalization factor. For example, in the leading order the anomalous dimension matrix is (see \cite{LN4} or \cite{Kotikov:2002ab}):
\begin{eqnarray}
\gamma^{(0)}_{{gg}}&=&-4S_1(j)+\frac{4}{j-1}-\frac{4}{j}+\frac{4}{j+1}-\frac{4}{j+2}\,,\qquad \
\gamma^{(0)}_{{\lambda g}} \, =\, \frac{8}{j}-\frac{16}{j+1}+\frac{16}{j+2}\,,
\nonumber \\[0mm]
\gamma^{(0)}_{{\phi g}}&=&\frac{12}{j+1}-\frac{12}{j+2}\,,\qquad
\gamma^{(0)}_{{g\lambda}} \, =\, \frac{4}{j-1}-\frac{4}{j}+\frac{2}{j+1}\,,\qquad
\gamma^{(0)}_{{\lambda\phi}} \, =\, \frac{8}{j}\,,\qquad
\gamma^{(0)}_{{\phi\lambda}} \, =\, \frac{6}{j+1}\,,
\nonumber \\[0mm]
\gamma^{(0)}_{{\lambda\lambda}} &=&-4S_1(j)+\frac{4}{j}-\frac{4}{j+1}\,,\qquad \
\gamma^{(0)}_{{g\phi}} \, =\, \frac{4}{j-1}-\frac{4}{j}\,,\qquad \
\gamma^{(0)}_{{\phi \phi}} \, =\, -4 S_1(j)\,.
\label{LOAD}
\end{eqnarray}
Sandwiched the operator from 
Eq.~(\ref{mrop1j}) between the different states we obtain the following combinations of the corresponding anomalous dimensions
\begin{eqnarray}
\gamma_{gg}+      \gamma_{\lambda g}+     \gamma_{\phi g}     &\ =\ &\gamma_{\mathrm {uni}}^{(0)}(j),\label{Op1g}\\[1mm]
\gamma_{g\lambda}+\gamma_{\lambda\lambda}+\gamma_{\phi\lambda}&\ =\ &\gamma_{\mathrm {uni}}^{(0)}(j),\label{Op1l}\\[1mm]
\gamma_{g\phi}+   \gamma_{\lambda\phi}+   \gamma_{\phi\phi}   &\ =\ &\gamma_{\mathrm {uni}}^{(0)}(j),\label{Op1s}
\end{eqnarray}
which will have the same value $\HS_1(j-2)$.

Next, sandwiched the operator from 
Eq.~(\ref{mrop2j}) between the different states we obtain the following combinations of the corresponding anomalous dimensions
\begin{eqnarray}
                 \gamma_{gg}   -\frac{1}{2(j-1)}\,\gamma_{\lambda g}     +\frac{1}{3}\,\frac{j+1}{j-1}\,\gamma_{\phi g}&\ =\ &\gamma_{\mathrm {uni}}^{(0)}(j+2),\label{Op2g}\\[1mm]
-\,2\,(j-1)\,\gamma_{g\lambda}+                 \gamma_{\lambda\lambda}+\frac{2}{3}\,(j+1)\,          \gamma_{\phi\lambda}&\ =\ &\gamma_{\mathrm {uni}}^{(0)}(j+2),\label{Op2l}\\[1mm]
-\,3\,\frac{j-1}{j+1}\,\gamma_{g\phi}+\frac{3}{2(j+1)}\,\gamma_{\lambda\phi}   +                          \gamma_{\phi\phi}&\ =\ &\gamma_{\mathrm {uni}}^{(0)}(j+2),\label{Op2s}
\end{eqnarray}
where we normalize the diagonal elements to unity, which will give the same value $\HS_1(j)$.

The last, sandwiched the operator from 
Eq.~(\ref{mrop3j}) between the different states we obtain the following combinations of the corresponding anomalous dimensions
\begin{eqnarray}
\gamma_{gg}      -\frac{j+2}{j-1}\,\gamma_{\lambda g}+\frac{j+1}{j}\,\frac{j+2}{j-1}\,\gamma_{\phi g}&\ =\ &\gamma_{\mathrm {uni}}^{(0)}(j+4),\label{Op3g}\\[1mm]
-\,\frac{j-1}{j+2}\,\gamma_{g\lambda}+          \gamma_{\lambda\lambda}-\frac{j+1}{j}               \gamma_{\phi\lambda}&\ =\ &\gamma_{\mathrm {uni}}^{(0)}(j+4),\label{Op3l}\\[1mm]
\frac{j-1}{j+2}\,\frac{j}{j+1}\,\gamma_{g\phi}-\frac{j}{j+1}\,\gamma_{\lambda\phi}+                            \gamma_{\phi\phi}&\ =\ &\gamma_{\mathrm {uni}}^{(0)}(j+4),\label{Op3s}
\end{eqnarray}
where again the diagonal elements are normalized to unity. These combinations have the same value $\HS_1(j+2)$.

Eqs.~(\ref{Op1g})-(\ref{Op3s}) can considerable simplify the calculations of a higher moments for the universal anomalous dimension $\gamma_{\mathrm {uni}}(j)$.
Thus, if we want to calculate the eight moment of the universal anomalous dimension $\gamma_{\mathrm {uni}}(8)$ we can use one from the Eqs.~(\ref{Op3g})-(\ref{Op3s}) and we will need the operators of Eqs.~(\ref{ggs})-(\ref{phphs}) only with $j=4$. Moreover we can further simplify our calculations, because the computations of a different operators (\ref{ggs})-(\ref{phphs}) sandwiched between a different states may differ considerably (for example, the computation of the anomalous dimension of gauge field operator (\ref{ggs}) sandwiched between the gauge fields is much complicated with compare to all other cases).

From our experience the most simple operator is the gaugino operator (\ref{qqs}). So, we take the seconds lines in the above equations with the combinations of the anomalous dimensions, i.e. Eqs.~(\ref{Op1l}),~(\ref{Op2l}) and~(\ref{Op3l}):
\begin{eqnarray}
\gamma_{g\lambda}+\gamma_{\lambda\lambda}+\gamma_{\phi\lambda}&\ =\ &\gamma_{\mathrm {uni}}^{(0)}(j),\label{Op1lj}\\[1mm]
-\,2\,(j-1)\,\gamma_{g\lambda}+                 \gamma_{\lambda\lambda}+\frac{2}{3}\,(j+1)\,          \gamma_{\phi\lambda}&\ =\ &\gamma_{\mathrm {uni}}^{(0)}(j+2),\label{Op2lj}\\[1mm]
-\,\frac{j-1}{j+2}\,\gamma_{g\lambda}+          \gamma_{\lambda\lambda}-\frac{j+1}{j}               \gamma_{\phi\lambda}&\ =\ &\gamma_{\mathrm {uni}}^{(0)}(j+4).\label{Op3lj}
\end{eqnarray}
We have three equations and six unknowns in a general case. For any particular values of $j$, for example $j=2$, if we fix $\gamma_{\mathrm {uni}}^{(0)}(j=2)$, $\gamma_{\mathrm {uni}}^{(0)}(j+2=4)$ from previous computations and calculate $\gamma_{\lambda\lambda}(j=2)$ then we can obtain the result for $\gamma_{\mathrm {uni}}^{(0)}(j+4=6)$ without any calculation of $\gamma_{g\lambda}$ and $\gamma_{\phi\lambda}$. Then we can proceed with the following $j=4$ to obtain $\gamma_{\mathrm {uni}}^{(0)}(8)$ and so on. Thus we need to calculate the anomalous dimension of operator (\ref{qqs}) with $j=4$ sandwiched between the gaugino state.

In alternative and even in a more simple way we can calculate the anomalous dimension $\gamma_{\lambda\phi}$ of operator (\ref{qqs}) with $j=4$ sandwiched between the scalar state. In this case we need the following three equations with $j=4$
\begin{eqnarray}
\gamma_{g\phi}+   \gamma_{\lambda\phi}+   \gamma_{\phi\phi}   &\ =\ &\gamma_{\mathrm {uni}}^{(0)}(j),\label{Op1sj}\\[1mm]
-\,3\,\frac{j-1}{j+1}\,\gamma_{g\phi}+\frac{3}{2(j+1)}\,\gamma_{\lambda\phi}   +                   \gamma_{\phi\phi}&\ =\ &\gamma_{\mathrm {uni}}^{(0)}(j+2),\label{Op2sj}\\[1mm]
\frac{j-1}{j+2}\,\frac{j}{j+1}\,\gamma_{g\phi}-\frac{j}{j+1}\,\gamma_{\lambda\phi}+                \gamma_{\phi\phi}&\ =\ &\gamma_{\mathrm {uni}}^{(0)}(j+4),\label{Op3sj}
\end{eqnarray}
which should be solved with known values of $\gamma_{\mathrm {uni}}^{(0)}(4)$, $\gamma_{\mathrm {uni}}^{(0)}(6)$ and $\gamma_{\lambda\phi}(4)$.

If we go to the four loops we can do the same as in the leading order for the contribution to the universal anomalous dimension, which appears for the first time at this order because there are no additional contributions from the renormalizations (see, however, below). We already used this property for our previous four-loop calculations~\cite{Velizhanin:2008pc,Velizhanin:2010ey}.

Thus, according to Eqs.~(\ref{Op1lj})-(\ref{Op3lj}) or Eqs.~(\ref{Op1sj})-(\ref{Op3sj}) we need to calculate the anomalous dimension $\gamma_{\lambda\lambda}(4)$ or $\gamma_{\lambda\phi}(4)$ and look only at the pole with quartic Casimir operator $d_{44} =N^2(N^2+36)/24$. This can be done with our program BAMBA based on the algorithm of Laporta~\cite{Laporta:2001dd}, which we used in our previous calculations. Laporta's algorithm allows to reduce with the help of an integration-by-parts identities all Feynman integrals to the finite (and small) numbers of a masters-integrals.
We used, following Refs.~\cite{Vladimirov:1979zm} (see also Refs.~\cite{Misiak:1994zw,Chetyrkin:1997fm,Czakon:2004bu}  for details), the infra-red rearrangement (IRR) procedure which reduce a propagator-type diagrams to a fully massive tadpole diagrams,
but produce some new difficulties related with an appearance of a non-gauge-invariant operators (see below).
For the reduction of the newly appeared Feynman integrals with a higher powers of denominators and numerators we considerably improved our {\texttt{MATHEMATICA}} code BAMBA.

The calculations of diagrams were performed with FORM~\cite{Vermaseren:2000nd}, using FORM package COLOR~\cite{vanRitbergen:1998pn} for evaluation of the color traces and with the Feynman rules for $\cN=4$ SYM theory from Ref.~\cite{Gliozzi:1976qd} and from Ref.~\cite{Bierenbaum:2009mv} for the operator $\mathcal{O}_{\mu _{1},...,\mu _{\M}}^{\lambda }$ with the different number of the gauge fields and with projectors from Refs.~\cite{Bierenbaum:2009mv,Gracey:2006zr}. For the dealing with a huge number of diagrams we use a program DIANA~\cite{Tentyukov:1999is}, which call QGRAF~\cite{Nogueira:1991ex} to generate all diagrams.

\begin{figure}
\begin{center}
  \includegraphics[width=90mm]{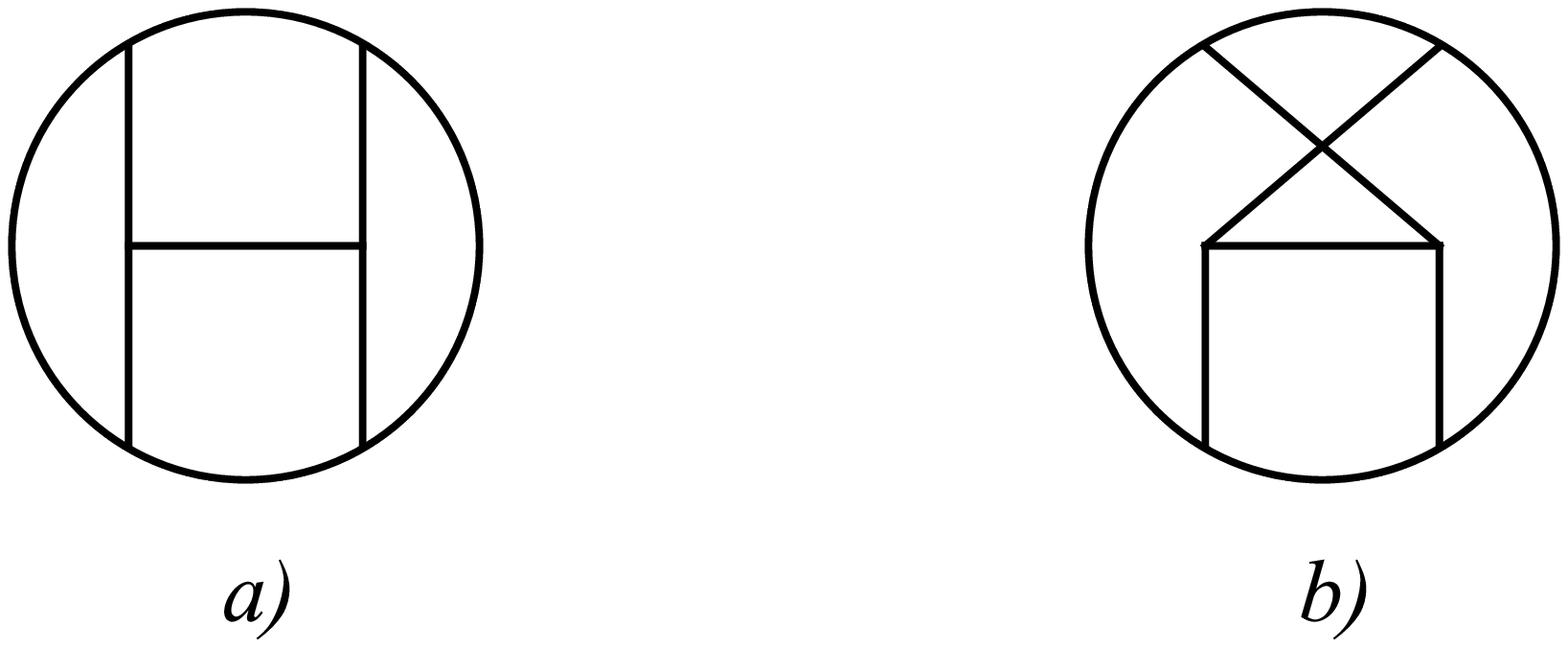}
    \caption {\textsf {
Basic parent four-loop planar
and non-planar
tadpole topologies.
}}\label{L4P}
\end{center}
\end{figure}

Practically, all computations are divided into three parts: calculation of the planar-based diagrams, calculation of the non-planar-based diagrams and a renormalization. At four loops in the method, which we used, there are two basic parent topologies for the fully massive tadpoles: planar and non-planar (see Fig.~\ref{L4P}).
All other topologies (see Ref.~\cite{Czakon:2004bu}) can be obtained by canceling one and more lines inside the parent topologies.

We use a slightly different codes for the calculations of the diagrams with these two different topologies. Moreover, this allow to have an additional test of the correctness of calculations, because separately the results for all diagrams with planar and non-planar parent topologies will have a complicated structure, i.e. include a higher poles and a lot of special numbers like $\zeta_i$ and other, while in the sum the final result should contain only the first poles with $\zeta_5$ and  $\zeta_3$ (probably, $\zeta_4$). Indeed, we obtain for $\hat\gamma_{\lambda\lambda}$ and $\hat\gamma_{\lambda\phi}$ anomalous dimensions (which are the parts of $\gamma_{\lambda\lambda}$ and $\gamma_{\lambda\phi}$) the following results:
\begin{eqnarray}
\hat\gamma_{\lambda\lambda}^{(3),{\mathrm{np}}}(4) & = &
\left(
\frac{12901}{225}
- \frac{68}{3}\,{\mathsf{S}}_2
+ \frac{417091}{2430}\,\zt
- \frac{9962}{15}\,\zf
\right)
\,
\frac{48\,g^8}{N_c^2}\,,
\label{gllM4op}\\[2mm]
\hat\gamma_{\lambda\phi}^{(3),{\mathrm{np}}}(4) & = &
\left(
\frac{809357}{1728}
+ \frac{91943}{450}\,  {\mathsf{S}}_2
+ \frac{742594303}{466560}\, \zt
- \frac{23072}{9}\, \zf
\right)
\,  \frac{48\,g^8}{N_c^2}\,,
\label{glsM4op}
\end{eqnarray}
where ${\mathsf{S}}_2$ is a special number like $\zeta_i$ (see Ref.~\cite{Czakon:2004bu}), which should absent in the final result.

As was mentioned above, because the non-planar contribution appears for the first time at the four loops then it should not have any renormalization. However, this is correct only for the gauge-invariant operators. In our method of calculation (see Ref.~\cite{Misiak:1994zw,Czakon:2004bu}) we introduce an auxiliary mass in the denominator of all propagators in Feynman integrals to reduce a propagator-type scalar diagrams to a tadpole-type diagrams, what considerably simplify the reduction to the master-integrals. In the final result this auxiliary mass will absent, but it can produce a {\it gauge-non-invariant} operators. In the calculation of a wave function renormalization constants such new gauge-non-invariant operator is a mass of the vector field ${\mathcal A}^{\mu}$
\begin{equation}
m^2 {\mathcal A}^{\mu}{\mathcal A}_{\mu}\,,
\label{mAA}
\end{equation}
which should be renormalized in addition to the usual renormalization and will give contribution under the renormalization.

\begin{figure}
\begin{center}
\includegraphics[width=140mm]{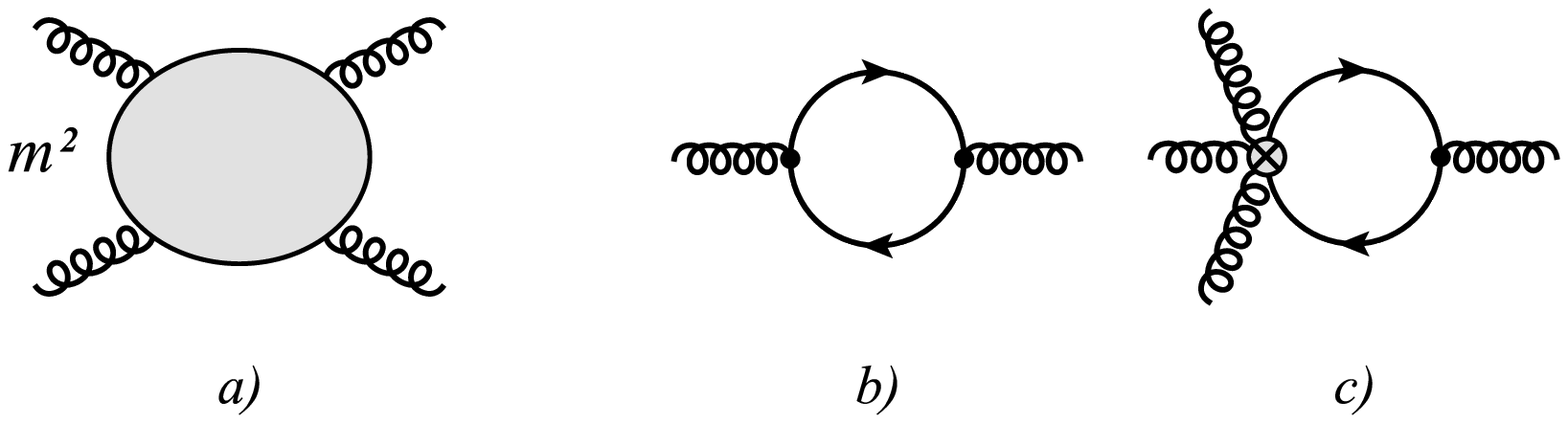}
  \caption {\textsf {
Gauge-non-invariant operator  $(a)$, relevant for our calculations.
The examples of the diagrams, which contribute to the renormalization of the mass $m^2$ of vector field $(b)$ and the proportional to $m^2$ four vector-field operator $(c)$.
}}\label{Fig:Op}
\end{center}
\end{figure}

In our case we can construct a several such operators, which are proportional to $m^2$, but we have found, that for the non-planar contribution only following one is relevant
\begin{equation}
m^2 {\mathcal A}^{\mu_1}{\mathcal A}^{\mu_2}{\mathcal A}^{\mu_3}{\mathcal A}^{\mu_4}\,,
\label{mAAAA}
\end{equation}
which presented on the Fig.~\ref{Fig:Op}~$a)$
and it should contain the fermion operator ${\mathcal O}^\lambda_{\mu_1,\mu_2,\mu_3,\mu_4}$ from Eqs.~(\ref{qqs}) inside a blob. Its appearance is very similar to the appearance of the vector field mass~(\ref{mAA}), i.e. two diagrams in Fig.~\ref{Fig:Op}~$b)$ and Fig.~\ref{Fig:Op}~$c)$  have the same divergencies, because the fermion operator with five line (three external gauge fields) in Fig.~\ref{Fig:Op}~$c)$  don't change the internal structure of diagram in Fig.~\ref{Fig:Op}~$b)$  (but there are a contributions from another diagrams, of course).

\begin{figure}
\begin{center}
\includegraphics[width=150mm]{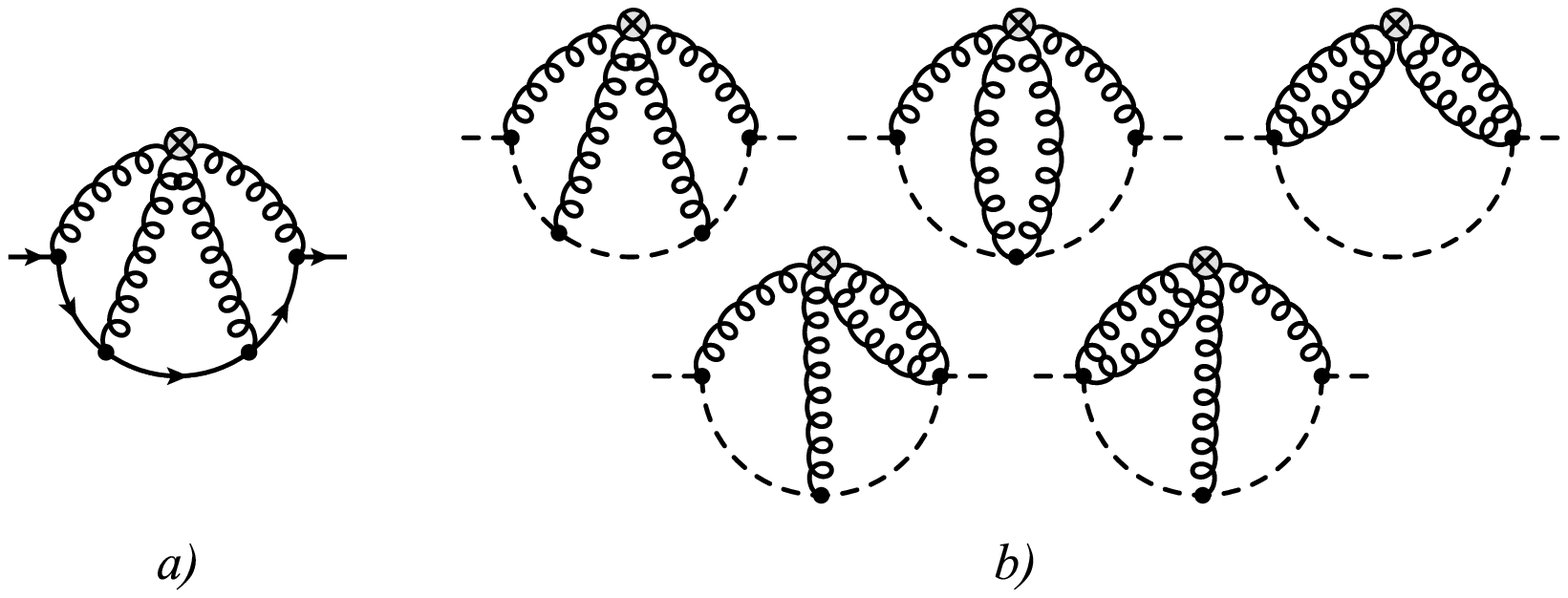}\hspace*{10mm}
  \caption {\textsf {
The diagrams with counterterm of the gauge-non-invariant operators Fig.~\ref{Fig:Op} $a)$ which give non-zero contributions to the anomalous dimensions
$\gamma_{\lambda\lambda}$ and $\gamma_{\lambda\phi}$.
}}\label{Fig:Diag}
\end{center}
\end{figure}

We computed the one-loop renormalization for this operator keep all Lorentz and color indices. Then, we insert it into corresponding three-loop diagrams. In the case of $\gamma_{\lambda\lambda}$ anomalous dimension there is only one diagram Fig.~\ref{Fig:Diag} $a)$, which gives non-zero contribution. In the case of $\gamma_{\lambda\phi}$ anomalous dimension there are five such diagrams Fig.~\ref{Fig:Diag} $b)$.

So, we obtain the following contributions from the gauge-non-invariant operator Fig.~\ref{Fig:Op}~$a)$ to $\gamma_{\lambda\lambda}$ and $\gamma_{\lambda\phi}$ anomalous dimensions:
\begin{eqnarray}
\gamma_{\lambda\lambda}^{(3),{{\mathrm{np}},\mathrm{ren}}}(4) & = &
\bigg(
- \frac{4}{3}
+ \frac{68}{3}\,{\mathsf{S}}_2
+ \frac{3781}{486}\,\zt
\bigg)\,
 \frac{48\,g^8}{N_c^2}\,,
\label{gllM4r}\\[2mm]
\gamma_{\lambda\phi}^{(3),{{\mathrm{np}},\mathrm{ren}}}(4) & = &
\bigg(
\frac{71999}{8640}
- \frac{91943}{450}\,{\mathsf{S}}_2
- \frac{35081983}{466560}\,\zt
\bigg)\,
  \frac{48\,g^8}{N_c^2}\,.
\label{glsM4r}
\end{eqnarray}

Collecting correspondingly Eqs.~(\ref{gllM4op}) and~(\ref{gllM4r}) and Eqs.~(\ref{glsM4op}) and~(\ref{glsM4r}) we obtain
\begin{eqnarray}
\gamma_{\lambda\lambda}^{(3),{{\mathrm{np}}}}(4) & = &
\bigg(
\frac{12601}{225}
+ \frac{8074}{45}\,\zt
- \frac{9962}{15}\,\zf
\bigg)\,
  \frac{48\,g^8}{N_c^2}\,,
\label{gllM4}\\
\gamma_{\lambda\phi}^{(3),{{\mathrm{np}}}}(4) & = &
\bigg(
\frac{21452}{45}
+ \frac{13648}{9}\,\zt
- \frac{23072}{9}\,\zf
\bigg)\,
  \frac{48\,g^8}{N_c^2}\,.
\label{glsM4}
\end{eqnarray}
In both cases ${\mathsf{S}}_2$ is exactly canceled, while other parts are slightly changed, except of $\zf$ contribution.

Substitute the obtained results (\ref{gllM4}) and (\ref{glsM4}) correspondingly into Eqs.~(\ref{Op1lj})-(\ref{Op3lj}) or Eqs.~(\ref{Op1sj})-(\ref{Op3sj}) we obtain the same result for the non-planar (or color subleading) contribution to the eight moment of the universal anomalous dimension of twist-2 operators:
\begin{equation}
\gamma_{\mathrm {uni,np}}^{(3)}(8)\ =\
\frac{49}{600}\, \Big(1357 + 434\,\zt - 11760\,\zf\Big)\,.
\label{guniM8}
\end{equation}
Note again, that we obtain the same result from two different ways: from $\gamma_{\lambda\lambda}^{(3),{{\mathrm{np}}}}(4)$ and Eqs.~(\ref{Op1lj})-(\ref{Op3lj}) and from $\gamma_{\lambda\phi}^{(3),{{\mathrm{np}}}}(4)$ and Eqs.~(\ref{Op1sj})-(\ref{Op3sj}). This gives an additional confirmation of the correctness of our result.

From Eqs.~(\ref{guniM4}), (\ref{guniM6}) and (\ref{guniM8}) we can see one interesting future: an overall factor is proportional to $S_1(j-2)$, i.e. the general expression for the color subleading contribution to the universal anomalous dimension of twist-2 operators may have the following form
\begin{equation}
\gamma_{\mathrm {uni,np}}^{(3)}(j)\ =\ S_1(j-2) \times\Big[\ldots,S_{i_1,i_2,\ldots}(j),\ldots,\zeta_i\,,\ldots\Big]\,,
\label{S1fM8}
\end{equation}
where brackets contain all other possible terms.

The $\zf$ part of the result~(\ref{guniM8}) coincide with the prediction from the general expression for the $\zf$ contribution to the non-planar four-loop universal anomalous dimension, suggested in Ref.~\cite{Velizhanin:2009gv}:
\begin{equation}\label{HSZ5ResS1}
\gamma_{{\mathrm{uni}},{\mathrm{np}},\,\zf}^{(3)}(\M)\ =\ -\,160\,\HS_{1}^2(\M-2)
\end{equation}
with
\begin{equation}\label{HSZ5ResALL}
\gamma_{{\mathrm{uni}},{\mathrm{np}}}^{(3)}(\M)=\left(\gamma_{{\mathrm{uni}},{\mathrm{np}},\,\zf}^{(3)}(\M)\,\zf+\gamma_{{\mathrm{uni}},{\mathrm{np}},\,\zt}^{(3)}(\M)\,\zt
+\gamma_{{\mathrm{uni}},{\mathrm{np}},{\mathrm{rational}}}^{(3)}(\M)\right)\frac{48}{N_c^2}\,.
\end{equation}

In Ref.~\cite{Velizhanin:2010ey} we also suggested a several variants for the $\zt$ contribution, but we have checked, that the $\zt$ part of~(\ref{guniM8}) does not coincide with any from these predictions. Moreover, we tried to reconstruct the $\zt$ contribution and/or the general result for the non-planar contribution to the four-loop universal anomalous dimension of twist-2 operators with the help of the methods, developed in our previous work~\cite{Velizhanin:2013vla} (see also Refs.~\cite{Velizhanin:2010cm,Velizhanin:2012nm}), using all possible basis from the harmonic sums, but unfortunately we have not found any reasonable solution.

\section{Discussion}

In this paper we present the result (\ref{guniM8}) for the eight moment of the non-planar contribution to the four-loop universal anomalous dimension of twist-2 operators in $\cN=4$ SYM theory, which was computed by the {\it full direct diagrammatic} calculations. The obtained result confirmed the general form of $\zf$ contribution~(\ref{HSZ5ResS1}) suggested by us in Ref.~\cite{Velizhanin:2009gv}, while did not allow reconstruct a general form of $\zt$ and rational parts due to breaking of some properties, which we used for the reconstructions in the planar limit.
We are going to continue to study of a suitable basis and the calculations of the higher even moments with the help of Eqs.~(\ref{Op1lj})-(\ref{Op3sj}), or odd moments for other relevant operators.
Nevertheless, we want to stress again two interesting futures of our results:
\begin{enumerate}
  \item The result for Konishi is very simple and contains {\it only} $\zf$ contribution:
\begin{equation}
\gamma^{\mathrm {4-loop,\,np}}_{\mathrm{Konishi}}\ =\ \gamma_{\mathrm {uni,np}}^{(3)}(j\!=\!4)\ =\ -\,\frac{17280}{N_c^2}\
\zf\,g^8,\qquad g^2\ =\ \frac{g^2_{YM}\,N_c}{(4\,\pi)^2}\,. \label{resnpKK}
\end{equation}
That is, the result contains {\it only} the leading transcendental contribution.
If this property will hold at the higher loops, then, probably, the general result for Konishi at any order may be obtained similar to~\cite{Leurent:2013mr}.
Moreover, looks rather reasonable, that the $\gamma_{\mathrm {uni,np}}^{(3)}(j)$ for $j=3$ contains also {\it only} the leading transcendental contribution, that is
\begin{equation}
\gamma_{\mathrm {uni,np}}^{(3)}(j\!=\!3)\ =\ -\frac{7680}{N_c^2}\ \zf\,g^8\,.
\label{resnpOIS}
\end{equation}
As it was conjectured in~\cite{Gunnesson:2009nn} the anomalous dimension of the one impurity state in the $\beta\!=\!\!\frac12-\!\!$~deformed theory coincides with the anomalous dimension of the $Z{\mathcal D}Z$-operator, or, in our notation, with $\gamma_{\mathrm {uni}}(j\!=\!3)$. It would be interesting to compare our prediction~(\ref{resnpOIS}) with the corresponding result in the $\beta\!=\!\!\frac12-\!\!$~deformed theory.
  \item It seems, that the general result for the color subleading contribution to the four-loop universal anomalous dimension of twist-2 operators in $\cN=4$ SYM theory have the following form:
\begin{equation}
\gamma_{\mathrm {uni,np}}^{(3)}(j)\ =\ S_1(j-2) \times\Big[\ldots,S_{i_1,i_2,\ldots}(j-2),\ldots,\zeta_i\,,\ldots\Big]\,,
\label{S1fM8A}
\end{equation}
i.e., it is proportional to $S_1(j-2)$. In the planar limit such part of the general result is responsible for a scaling function~\cite{Beisert:2006ez}. So, the non-planarity may be related directly with the scaling function.
\end{enumerate}

 \subsection*{Acknowledgments}
I would like to thank L.N. Lipatov, A.I. Onishchenko, A.V. Smirnov, V.A. Smirnov and M.~Staudacher for useful discussions.
This research is supported by a Marie Curie International Incoming Fellowship within the 7th European Community Framework Programme,
grant number PIIF-GA-2012-331484,
by DFG, SFB 647 \emph{Raum -- Zeit -- Materie. Analytische und Geometrische Strukturen}
and by RFBR grants 12-02-00412-a, 13-02-01246-a.

\end{document}